\documentclass[11pt,a4paper]{article}
\usepackage{jheppub}
\usepackage{mathrsfs}
\usepackage{amsfonts}
\usepackage{mathtools}
\usepackage{setspace}
\usepackage{cellspace}
\usepackage{amsmath,amssymb,bm}
\usepackage[colorlinks=true,linkcolor=blue]{hyperref}
\usepackage{xcolor}
\usepackage{epsfig}
\usepackage{slashed}
\usepackage{caption}
\usepackage{hhline,multirow,tabularx}  
\usepackage{dcolumn}    
\usepackage{url}        
\usepackage{braket}     
\usepackage{pifont}
\title{Generalized parton distributions through universal moment parameterization: non-zero skewness case}

\author[a,d]{Yuxun~Guo}
\author[a]{, Xiangdong~Ji}
\author[b]{, M.~Gabriel~Santiago}
\author[c]{, Kyle~Shiells}
\author[a]{and Jinghong~Yang}

\affiliation[a]{Maryland Center for Fundamental Physics, Department of Physics, University of Maryland,\\ 4296 Stadium Dr., College Park, MD 20742, USA}
\affiliation[b]{Center for Nuclear Femtography, SURA,\\ 1201 New York Ave. NW, Washington, DC 20005, USA}
\affiliation[c]{Department of Physics and Astronomy, University of Manitoba,\\ Allen Building, Winnipeg, MB Canada R3T 2N2}
\affiliation[d]{Nuclear Science Division, Lawrence Berkeley National Laboratory,\\ Berkeley, CA 94720, USA}
\emailAdd{yuxunguo@umd.edu}
\emailAdd{xji@umd.edu}
\emailAdd{gsantiago@sura.org}
\emailAdd{kylelebaron.shiells@umanitoba.ca}
\emailAdd{yangjh@umd.edu}

\abstract{We present the first global analysis of generalized parton distributions (GPDs) combing lattice quantum chromodynamics (QCD) calculations and experiment measurements including global parton distribution functions (PDFs), form factors (FFs) and deeply virtual Compton scattering (DVCS) measurements. Following the previous work where we parameterize GPDs in terms of their moments, we extend the framework to allow for the global analysis at non-zero skewness. Together with the constraints at zero skewness, we fit GPDs to global DVCS measurements from both the recent JLab and the earlier Hadron-Electron Ring Accelerator (HERA) experiments with two active quark flavors and leading order QCD evolution. With certain choices of empirical constraints, both sea and valence quark distributions are extracted with the combined inputs, and we present the quark distributions in the proton correspondingly. We also discuss how to extend the framework to accommodate more off-forward constraints beyond the small $\xi$ expansion, especially the lattice calculated GPDs.}
\date{\today}
\begin{document}
\maketitle

\section{Introduction}

Over the past two decades, there has been growing interest in the higher-dimensional structures of the nucleon in the studies of the non-perturbative strong interactions described by quantum chromodynamics (QCD).
Consequently, generalized parton distributions (GPDs)~\cite{Muller:1994ses, Ji:1996ek, Ji:1998pc} that unify the elastic form factors (FFs) and Parton Distributions Functions (PDFs) into a single set of 3-dimensional (3D) functions and contain important information about the nucleon mass, angular momentum and mechanical properties~\cite{Ji:1994av, Ji:1996ek, Polyakov:2002yz} have gained increasing attention. While it is believed that GPDs can be experimentally accessed by exclusive productions of particles off nucleons such as deeply virtual Compton scattering (DVCS) \cite{Ji:1996nm} and deeply virtual meson production (DVMP)~\cite{Radyushkin:1996ru,Collins:1996fb}, obtaining them and the corresponding nucleon 3D structures from experimental data has remained a dream. On the one hand, persisting efforts are needed for an adequate amount of data as required for the extraction of such high-dimensional quantities. On the other hand, the extraction of the critical information from experimental data has also been challenging.

Several recent breakthroughs have made this possible. Large data sets of exclusive measurements are generated at Jefferson Lab (JLab)~\cite{CLAS:2018ddh,CLAS:2021gwi,Georges:2017xjy,JeffersonLabHallA:2022pnx}, and more will come from the planned Electron-Ion Collider (EIC)~\cite{Accardi:2012qut, AbdulKhalek:2021gbh} at Brookhaven National Lab (BNL). Additionally, novel developments in lattice QCD allow the access to the nucleon structures from first principle calculation~\cite{Ji:2013dva,Ji:2020ect,Alexandrou:2020zbe,Lin:2021brq}, providing information almost impossible to get with current and future experiments.
This recent progress, together with the perennial efforts on the global extraction of PDFs~\cite{Hou:2019efy,NNPDF:2021njg,Ethier:2017zbq}, FFs~\cite{Ye:2017gyb} as well as lattice calculation of generalized form factors~\cite{Alexandrou:2021jok,Hasan:2017wwt,Shintani:2018ozy,Jang:2018djx}, have pushed the study of nucleon 3D structures to a new stage, which, on the other hand, requires extra works to put all the above inputs together through global analysis to obtain the state-of-art nucleon 3D structures with GPDs.

Therefore, we are in need of a global analysis program of GPDs with both experimental data and lattice computation. Lots of efforts are put into parameterizing and accessing GPDs from various inputs in the literature~\cite{Polyakov:2002wz,Guidal:2004nd,Goloskokov:2005sd,Mueller:2005ed,Kumericki:2009uq,Goldstein:2010gu,Gonzalez-Hernandez:2012xap,Kriesten:2021sqc,Hashamipour:2021kes}. However, a framework that could combine experimental data and lattice computation still seems to be lacking. In the previous work~\cite{Guo:2022upw}, we proposed the GPDs through universal moment parameterization (GUMP) program for such a purpose and studied the zero-skewness case as an illustrative example. In this work, we follow the previous setup and extend it to non-zero skewness, which allows us to also include the exclusive measurements in the global analysis. With such a program, we perform the global analysis of GPDs that includes global PDFs~\cite{Hou:2019efy,NNPDF:2021njg,Ethier:2017zbq}, FFs~\cite{Ye:2017gyb} and DVCS measurements from both JLab~\cite{CLAS:2018ddh,CLAS:2021gwi,Georges:2017xjy,JeffersonLabHallA:2022pnx} and previous H1~\cite{H1:2009wnw} experiments at Hadron-Electron Ring Accelerator (HERA) along with the relevant lattice calculations related to GPDs~\cite{Alexandrou:2021jok,Alexandrou:2020zbe} for the first time. 

The organization of the paper is as follows. In sec. \ref{sec:setup}, we introduce the setup for the GUMP program and discuss how to build the 3D GPDs with a handleable set of parameters. In sec. \ref{sec:fit}, we discuss more details of the fit and present the extracted Compton form factors (CFFs) and GPDs, where we also discuss how to extend the present framework to accommodate more future inputs. In the end, we conclude in sec. \ref{sec:conclude}.

\section{Conventions and fitting procedure}

\label{sec:setup}

In the previous work~\cite{Guo:2022upw}, we discussed how GPDs can be parameterized in terms of their moments based on the technique that has been systematically developed in ref.~\cite{Mueller:2005ed} and used for instance in ref.~\cite{Kumericki:2009uq}. In this section, we will discuss how we apply such a framework to the GPDs at non-zero skewness, including GPDs of various different species and flavors.

\subsection{Convention and decomposition of GPD}

We start by considering the four leading-twist GPDs $H,E,\widetilde{H}$ and $\widetilde{E}$, where each of them has different flavors $u,d$ and $g$ corresponding to the up and down quarks as well as the gluon. Although the strange quark might have sizable effects that should be considered for a more careful treatment, the flavor separation of GPDs is typically challenging due to the lack of a flavor-sensitive probe~\cite{Cuic:2020iwt}. Strange quark distributions will not be well constrained with just DVCS measurements, and thus we will leave this for the future work with more flavor-sensitive measurements.

For simplicity, the four different species of GPDs will be treated almost equally in this work. Except that the vector GPDs $H$ and $E$ satisfy slightly different scale evolution equations from the axial-vector ones $\widetilde{H}$ and $\widetilde{E}$~\cite{Kumericki:2007sa}, the same parameterization will be used for all four GPDs. Although GPDs or combinations of GPDs of different species $H,E,\widetilde{H}$ and $\widetilde{E}$ correspond to different helicity amplitudes and might have different behaviors accordingly, knowledge of different GPDs species is limited and thus implementing a different parameterization for them individually may not lead to any significant difference. Therefore, we collectively denote all GPDs as $F_{q,g}(x,\xi,t)$ where $F=\{H,E,\widetilde{H},\widetilde{E}\}$ and $q=\{u,d\}$, with $x$ the parton momentum fraction, $\xi$ the skewness parameter and $t$ the total momentum transfer squared and parameterize them in the same manner. We note that the skewness parameter $\xi$ generally takes the value from $-1$ to $1$, but it will be considered to be positive hereafter, as GPDs are commonly defined to be symmetric in $\xi$, subject to their parity symmetry.

An intriguing feature of the GPDs is that they consist of two regions with totally different physical interpretations. In the PDF-like region where $x>\xi$ or $x<-\xi$, GPDs resemble the PDFs which are interpreted as the amplitudes of emitting and reabsorbing a parton (quark, antiquark or gluon). On the other hand, in the distribution amplitude (DA)-like region where $-\xi<x<\xi$, they resemble the DAs instead and are interpreted as the amplitudes of emitting/absorbing a parton-antiparton pair. \textit{Therefore, GPDs do not naturally have distinguishable quark and antiquark components, especially in the DA-like region.} This feature indicates that the parameterization of GPDs should be flexible in both the PDF and DA-like regions, since different physics are involved.

Motivated by such property, one can write GPDs as,
\begin{equation}
\label{eq:gpddecomp}
   F_{q}(x,\xi,t) \equiv F_{\hat{q}}(x,\xi,t)\mp F_{\bar{q}}(-x,\xi,t) + F_{q\bar q}(x,\xi,t)\ ,
\end{equation}
where the $\mp$ depends on the parity of the GPDs which takes $-$ for vector GPDs $H$ and $E$ and $+$ for axial vector GPDs $\widetilde{H}$ and $\widetilde{E}$. Here the subscript $\hat q$ stands for the quark distributions excluding antiquark, to be distinguished from the subscript $q$ for the quark flavor. The three terms above describe the amplitudes corresponding to the quark, antiquark, and quark-antiquark pair respectively. Accordingly, $F_{\hat{q}}(x,\xi,t)$ and $F_{\bar{q}}(x,\xi,t)$ have support $x>-\xi$, while the $F_{{q}\bar q}(x,\xi,t)$ has support $\xi>x>-\xi$. In the semi-forward limit where $\xi = 0 $ and $t$ generally non-zero, $F_{\hat{q}}(x,\xi,t)$ and $F_{\bar{q}}(x,\xi,t)$ reduce to the quark and antiquark $t$-dependent PDFs with support $x>0$, whereas $F_{{q}\bar q}(x,\xi,t)$ vanishes with its vanishing support. We note that GPDs can always be decomposed into such three parts that each term represents the behavior of GPDs in different regions respectively. The extra DA terms $F_{{q}\bar q}(x,\xi,t)$ allow extra flexibility for the parameterization of GPDs particularly in the DA-like region, as we will discuss with more details in the next section.

While the quark and antiquark GPDs $F_{\hat{q}}(x,\xi,t)$ and $F_{\bar{q}}(x,\xi,t)$ are the natural generalization of the quark and antiquark PDFs, the DA terms $F_{{q}\bar q}(x,\xi,t)$, also known as the $D$-terms in some other contexts~\cite{Polyakov:1999gs}, only exists in the off-forward case. Such terms live in the DA-like region only, and they will be the dominant effects in the large $\xi$ limit $\xi \to 1$ or the asymptotic limit where renormalization scale $\mu \to\infty$~\cite{Goeke:2001tz,Diehl:2003ny}. Since the DA terms vanish at the crossover line $x=\pm \xi$, they can always be expressed with a set of complete polynomials like Gegenbauer polynomials, which is exactly the conformal moment expansion. Therefore, each of the three terms can be expressed in terms of their conformal moments and satisfies the polynomiality condition. 

In this work, we will focus on the quark and antiquark GPDs $F_{\hat{q}}(x,\xi,t)$ and $F_{\bar{q}}(x,\xi,t)$, whereas the extra DA terms $F_{q\bar q}(x,\xi,t)$ will not be put into the global analysis for two reasons. First, as discussed in the previous work, we will consider the high energy limit where $\xi \to 0$ ideally and the DA-like region disappears in such a limit. Second, since the DA terms vanish at the crossover line $x=\pm\xi$, they are hard to extract with measurements of DVCS or similar processes which provide effectively just CFFs --- the DA terms contribute to the real part of the CFFs only, corresponding to the subtraction terms in the dispersion relations of CFFs~\cite{Kumericki:2009uq}. Consequently, the behaviors of GPDs in the DA-like region given by the quark and antiquark GPDs $F_{\hat{q}}(x,\xi,t)$ and $F_{\bar{q}}(-x,\xi,t)$ will be ambiguous, since one can add extra DA terms without affecting the CFFs too much in the small $\xi$ limit. This will be discussed with more details in the next section.

We note that another notation of quark and antiquark GPDs has been used in the literature, see for instance ref.~\cite{Muller:2014wxa}
\begin{align}
\begin{split}
    F_{\hat{q}}(x>\xi,\xi,t)&=F_{q}(x,\xi,t) \ ,\\
    F_{\bar q}(x>\xi,\xi,t)&=\mp F_{q}(-x,\xi,t)\ ,
\end{split}
\end{align}
with the same convention for the $\mp$ sign. Since the quark and antiquark GPDs are defined with the same function $F_{q}(x,\xi,t)$ partitioned into two parts, the full quark GPD $F_{q}(x,\xi,t)$ cannot be written as the sum of them $F_{q}(x,\xi,t) \not = F_{\hat{q}}(x,\xi,t)\mp F_{\bar{q}}(-x,\xi,t)$. Due to the loss of linearity in such a GPD decomposition, which will also break the polynomiality condition for quark and antiquark GPDs respectively, this definition will not be used in this work.\footnote{Although this decomposition is written in the previous work \cite{Guo:2022upw}, the shift of definition will not cause inconsistency as it was in the zero-skewness limit where the DA-like region does not exist.}

To summarize, we will consider quark and antiquark GPDs $F_{\hat q}(x,\xi,t)$, $F_{\bar q}(x,\xi,t)$ as well as the gluon GPDs $F_{g}(x,\xi,t)$ in this work with four twist-two GPDs $H,E,\widetilde{H}$ and $\widetilde{E}$, forming a basis with 20 GPDs. Conventionally, we also rewrite the basis by defining the valence combination: $F_{q_V}(x,\xi,t)\equiv F_{\hat q}(x,\xi,t)-F_{\bar q}(x,\xi,t)$, and thus the basis can be written with $F_{q_V}(x,\xi,t)$, $F_{\bar q}(x,\xi,t)$ and $F_{g}(x,\xi,t)$ equivalently, analogous to the basis commonly chosen for PDFs global analysis in the literature~\cite{Hou:2019efy,NNPDF:2021njg,Ethier:2017zbq}.

\subsection{Parameterization of moments}

\label{subsec:momentpara}

After introducing the GPDs necessary for consideration, we briefly discuss the parameterization method for GPDs. We have introduced the GUMP parameterization in the previous work~\cite{Guo:2022upw} and more about the conformal moment representation can be found in refs. \cite{Mueller:2005ed, Kumericki:2009uq}. Generally, we express all GPDs $F(x,\xi,t)$ in terms of their conformal moments $\mathcal F_j(\xi,t)$ in the form of,
\begin{equation}
\label{eq:conformalsum}
  F(x,\xi,t) = \sum_{j=0}^{\infty} (-1)^j p_j(x,\xi) \mathcal {F}_{j}(\xi,t) \quad \text{for }|x|<\xi\ ,
\end{equation}
with $p_j(x,\xi)$ the known conformal wave functions. Therefore, all GPDs $F(x,\xi,t)$ are equivalently given by $\mathcal {F}_{j}(\xi,t)$. With the polynomiality condition of GPDs~\cite{Ji:1998pc}, the moments can be expressed as polynomials of $\xi$ of given order:
\begin{equation}
\mathcal F_j(\xi,t)=\sum_{k=0,\rm{ even}}^{j+1} \xi^{k} \mathcal{F}_{j,k}(t) \ ,
\end{equation}
and then these $\mathcal{F}_{j,k}(t)$ can be used to construct the GPD $F(x,\xi,t)$.

In the high energy limit, the skewness parameter $\xi = x_B/(2-x_B) + \mathcal{O}(Q^{-2})$ is small where the first few terms dominate, and we have~\cite{Kumericki:2009uq}
\begin{equation}
\mathcal F_j(\xi,t)= \mathcal{F}_{j,0}(t) + \xi^2 \mathcal{F}_{j,2}(t) + \mathcal{O}(\xi^4)\ ,
\end{equation}
where we implicitly assume that the $\mathcal{F}_{j,k}(t)$ are only non-zero for $j>k$ according to the polynomiality condition. Practically the moments $\mathcal{F}_{j,k}(t)$ can be written in terms of the shape governed by the Euler beta function $B$, the Regge trajectory for the $t$-dependence and the extra residual term $\beta(t)$:
\begin{equation}
\label{eq:gumpform}
    \mathcal F_{j,k}(t)= \sum_{i=1}^{i_{\rm{max}}}N_{i,k} B(j+1-\alpha_{i,k},1+\beta_{i,k})\frac{j+1-k-\alpha_{i,k}}{j+1-k-\alpha_{i,k}(t)} \beta(t)\ ,
\end{equation}
with which the GPDs are parameterized in terms of the free parameters in eq. (\ref{eq:gumpform}). In the simplest case, only one set of ansatz is needed, so we set $i_{\rm{max}}$ to be just $1$.

In the previous work, we simply set the residual function $\beta(t)$ to be 1 for the extraction of valence quark distributions in the zero-skewness case, since their algebraic decaying behaviors in $|t|$ can be parameterized well by the Regge trajectory. However, for the sea distributions, it has been observed that in high energy processes, the different cross-sections drop exponentially as the momentum transfer $|t|$ increases e.g., for DVCS \cite{H1:2007vrx,H1:2009wnw}, which differs from the power-law behavior indicated by the Regge theory. Therefore, we incorporate the extra exponential behavior into the $\beta(t)$ to set $\beta(t) = \exp(b t)$. Two $\beta(t)$s of different $b$ slope are embedded in the parameterization of the vector GPDs $H_{\bar q,g}$ and the axial vector GPDs $\widetilde{H}_{\bar q, g}$ respectively, with $q=\{u,d\}$ for sea quarks. We note that due to the lack of flavor-sensitive probe, we assume the same $b$ slope for the gluon and sea quarks of different flavor and also the same Regge slope $\alpha'$ for them which is fixed to be $\alpha'_{\bar q,g}=0.15$ from the pomeron trajectory~\cite{H1:2005dtp}.

Even with such simplification, the parameters are still too many to be fully determined from measurements. The lack of off-forward constraints makes it almost impossible to get the shape of GPDs at non-zero skewness, which is known as the deconvolution problem of GPDs~\cite{Bertone:2021yyz}, stating that the shape of GPDs can not be uniquely determined from CFFs. Therefore, extra empirical constraints are still needed for the extraction of GPDs. The simplest choice for the $\xi$-dependent terms is to set them to be proportional to the leading terms, as has been done in ref.~\cite{Kumericki:2009uq}:
\begin{equation}
\label{eq:Rdef}
    \mathcal F_{j,k}(t) = R_{k} \mathcal F_{j-k,0}(t)\ ,
\end{equation}
with $R_{k}$ the ratio between them. In this work, we have just one parameter $R_2$ for the $\xi^2$ terms for each GPD $F(x,\xi,t)$ that accounts for its extra $\xi$-dependence. 

Another difficulty in GPD parameterization is about the flavors and species of GPDs. While the extraction of one 3D function is already challenging, the simultaneous extraction of GPDs with multiple flavors and species is yet more difficult. Two of the four different species of GPDs, $E$ and $\widetilde{E}$, do not have corresponding forward PDFs, unlike $H$ and $\widetilde{H}$ which reduce to the PDF $f(x)$ and helicity PDF $\Delta f(x)$ respectively, adding extra difficulties to their extraction. Consequently, we have to reduce the number of parameters associated to these two GPDs due to the lack of constraints. To do so, we set the two valence quark distributions to be proportional ($E_{u_V} \propto E_{d_V}$ and $\widetilde{E}_{u_V} \propto \widetilde{E}_{d_V}$) and the sea quark as well as the gluon distributions to be proportional to the corresponding  $H$ and $\widetilde{H}$ GPDs. 

We note that these extra empirical constraints, both for the $\xi$-dependent terms and for the $E/\widetilde{E}$ GPDs, are added for purely practical purposes. As mentioned above, they simply represent the lack of information of GPDs in the off-forward region, which can and should be improved in the future with more inputs from both lattice calculations and experimental measurements. In table \ref{table:GUMPparameters}, we collect whether GPDs of different species and flavors are either parameterized independently with eq. (\ref{eq:gumpform}) or linked to the other GPDs with some empirical assumptions in this work.

\begin{table}[t]
    \centering
    \begin{tabular}{|Sc|Sc|Sc|Sc|}
    \hline
       GPDs species and flavors & Fully parameterized& GPDs linked to & \begin{tabular}{c} Proportional\\ constants \end{tabular}   \\ \hline
       $H_{u_V}$ and $\widetilde{H}_{u_V}$ & \ding{52} & - & - \\ \hline
       $E_{u_V}$ and $\widetilde{E}_{u_V}$ & \ding{52} & - & - \\ \hline
       $H_{d_V}$ and $\widetilde{H}_{d_V}$ & \ding{52} & - & - \\ \hline
       $E_{d_V}$ and $\widetilde{E}_{d_V}$ & \ding{56} &  $E_{u_V}$ and $\widetilde{E}_{u_V}$ &  $R^{E/\widetilde{E}}_{d_V}$ \\ \hline
       $H_{\bar u}$ and $\widetilde{H}_{\bar u}$ & \ding{52} & - & - \\ \hline
       $E_{\bar u}$ and $\widetilde{E}_{\bar u}$ & \ding{56}  & $H_{\bar u}$ and $\widetilde{H}_{\bar u}$ & $R^{E/\widetilde{E}}_{\rm{sea}}$\\ \hline
       $H_{\bar d}$ and $\widetilde{H}_{\bar d}$ & \ding{52} & - & - \\ \hline
       $E_{\bar d}$ and $\widetilde{E}_{\bar d}$ & \ding{56}  &  $H_{\bar d}$ and $\widetilde{H}_{\bar d}$& $R^{E/\widetilde{E}}_{\rm{sea}}$\\ \hline
       $H_{g}$ and $\widetilde{H}_{g}$ & \ding{52} & -& -  \\ \hline
       $E_{g}$ and $\widetilde{E}_{g}$ & \ding{56}  &  $H_{g}$ and $\widetilde{H}_{g}$& $R^{E/\widetilde{E}}_{\rm{sea}}$\\ \hline
    \end{tabular}
    \caption{A summary of how each GPDs with different species and flavors are parameterized respectively. Fully parameterized GPDs are expressed in terms of eq. (\ref{eq:gumpform}), whereas the other GPDs are linked to the fully parameterized GPDs with proportional constants.}
    \label{table:GUMPparameters}
\end{table}

Finally, we comment on the comparison of the GUMP program to the other GPD global analysis programs with conform moments framework, specifically the Kumeri\v{c}ki-M\"uller (KM) model and its extensions~\cite{Kumericki:2007sa,Kumericki:2009uq,Muller:2013jur,Mueller:2014hsa}. As clearly stated in the previous work~\cite{Guo:2022upw}, the GUMP program follows the general conformal moment parameterization of GPDs, and adopts many useful observations there to make the GUMP parameterization practical. On the other hand, this program focuses more on the extraction of all four leading-twist GPDs including both valence and sea quarks of different flavors with the combined inputs from both lattice calculations and experiments. Therefore, though in a similar spirit, the GUMP program allows a more comprehensive study of the GPDs, especially the valence parts which are effectively constrained by lattice calculations while less accessible from experiments.

\subsection{Inputs and fitting strategy}

With the above parameterization, the global analysis can be performed with adequate inputs to pin down all the parameters. In this subsection, we will discuss how we select inputs from all the available results for the global analysis. 

We start with the forward inputs where GPDs reduce to PDFs. While we could have various inclusive measurements as the inputs rather than the global PDFs extracted from a specific work to avoid the bias, there are several reasons we choose the extracted PDFs for the GPD global analysis.
First, the extraction of the PDFs themselves is considerably involved and requires dedicated works, especially for the simultaneous extraction of both unpolarized and polarized PDFs~\cite{Cocuzza:2022jye,Zhou:2022wzm}. Second, much less is known for the off-forward behaviors of GPDs compared to the forward ones. Consequently, a fine forward analysis can not be matched with an equivalent off-forward analysis, making it less urgent to improve the forward part of the analysis. Therefore, we avoid repeating the forward fittings that have been studied extensively by the PDF global analysis community and take their globally extracted PDFs as the inputs. More specially, we consider one of the recent analyses by the JAM collaboration ~\cite{Cocuzza:2022jye}, where both the unpolarized and polarized proton PDFs are extracted simultaneously. 

The off-forward inputs, on the other hand, consist of various constraints including globally extracted FFs~\cite{Ye:2017gyb}, lattice calculations~\cite{Alexandrou:2021jok,Alexandrou:2020zbe} and exclusive measurements, or the DVCS measurements~\cite{CLAS:2018ddh,CLAS:2021gwi,Georges:2017xjy,JeffersonLabHallA:2022pnx,H1:2009wnw} more specifically for this work. For the form factors, we take the globally extracted charge FFs~\cite{Ye:2017gyb} rather than fit to the elastic scattering data directly~\cite{Hashamipour:2021kes} for the same reasons as for the PDFs. Since the charge form factors are averaged over all flavors, we combine the charge form factors of both proton and neutron to obtain the charge form factors of up and down quarks respectively, assuming isospin symmetry and ignoring the contributions of strange and other heavy flavors.

The lattice inputs include both calculations of generalized form factors and $x$-dependence of GPDs~\cite{Alexandrou:2021jok,Alexandrou:2020zbe}. Putting the lattice QCD results together with other experimental measurements can be quite subtle, due to the systematic uncertainties of the lattice results that are hard to fully understand. Progress has been made in getting the systematic uncertainties under control and calculations from different groups are converging for certain quantities like the (axial) charge form factors, see for instance the review in ref.~\cite{Constantinou:2020hdm}, but more efforts are still in demand to get the other results consistent. Therefore, in this work we consider the lattice results from a single group only~\cite{Alexandrou:2021jok,Alexandrou:2020zbe} \footnote{Here the choice is made by considering only the variety of GPD related calculations and does not reflect a preference over the other lattice calculations.} to avoid the potential tension among lattice results from different groups. On the other hand, we adjust their weights in the global analysis by increasing the uncertainties of the results to account for the unknown systematic uncertainties.
We note that the calculation of the $x$-dependence of GPDs at non-zero skewness has been done in the literature~\cite{Alexandrou:2020zbe}, although the reliable regions with controlled systematic uncertainties get even more subtle due to the irregular behavior of GPDs at $x=\xi$. Thus, the inclusion of the $x$-dependence of GPDs at non-zero skewness in the global analysis will be left to the future work.

Last but not the least, we have the experimental exclusive measurements. In principle, the exclusive measurements can and should include as many processes as possible to get better constraints on the GPDs as well as to test the universality of GPDs. However, the main challenge for putting different processes together is the mismatch in the amount of data available. Two types of Compton scattering processes, DVCS and Time-like Compton Scattering (TCS)~\cite{Berger:2001xd}, have been measured at JLab. Much more DVCS data have been obtained than that of TCS, of which the first measurement was made just recently~\cite{CLAS:2021lky}. On the other hand, DVMP typically requires much larger virtuality than DVCS to suppress the unwanted contributions of transverse polarization, which are mostly accessible with colliders such as HERA~\cite{ZEUS:2005bhf,H1:2005dtp} and the future EIC~\cite{Accardi:2012qut, AbdulKhalek:2021gbh}. Therefore, we start by considering the DVCS measurements only, which has the broadest global kinematical coverage among these process that includes both the low $x_B$ region covered by H1~\cite{H1:2009wnw} at HERA and the medium $x_B$ region covered by various experiments at  JLab~\cite{CLAS:2018ddh,CLAS:2021gwi,Georges:2017xjy,JeffersonLabHallA:2022pnx}. It is worth noting that the DVCS process is mostly sensitive to the quark distributions whereas the gluon distributions are best obtained from meson production or other gluon-sensitive processes. Obtaining the gluonic distributions from them is indeed of high interest and importance, which will be carried out in a separate work.

With all the inputs above, it seems straightforward to simply put them together and perform the global analysis with the parameterization described before. However, it will not be quite practical with the large number of free parameters, which typically means extremely slow convergence in the search of best-fit parameters. Besides, the evaluation of GPDs, especially with scale evolution, is much more computationally intensive, compared to that of the PDFs for instance, making the global analysis even less handleable. Therefore, we split the global analysis into two steps by first performing the semi-forward fit at zero skewness $\xi =0$ (while the momentum transfer squared $t$ can still be non-zero) and then fit to the off-forward inputs with non-zero skewness $\xi \not= 0$. Apparently, the two-step fitting introduces bias that favors the semi-forward constraints over the off-forward constraints. However, given the fact that much more semi-forward constraints can be obtained from lattice and experiments than the off-forwards ones, it is a reasonable and practical assumption for such a large system of parameters.

\section{Global analysis and extracted GPDs}
\label{sec:fit}
In the previous section, we discussed the parameterization of GPDs as well as the inputs to constrain the parameters, so we would need to find the set of parameters that fit to the measurements best, for which we employ the \textit{iminuit} interface of \textit{Minuit2}~\cite{iminuit,James:1975dr} as the minimizer. In this section, we will present the results of the global analysis as well as the extracted CFFs and GPDs.

\subsection{Basics of the fit}

\begin{figure}[t]
\centering
\includegraphics[width=\textwidth]{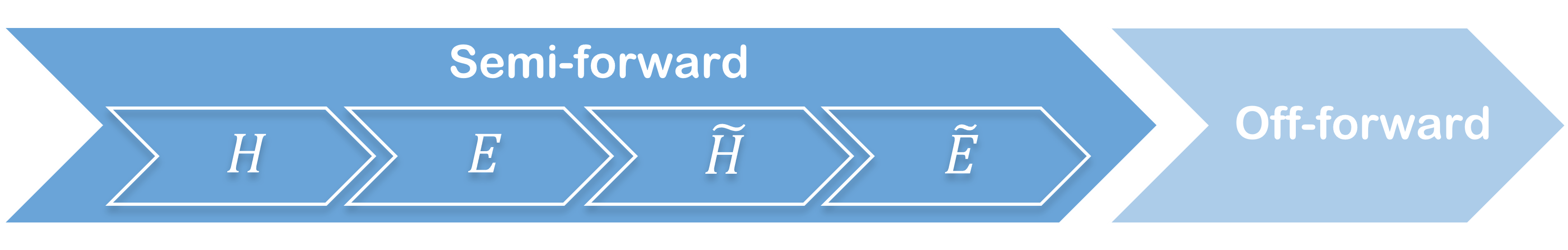}
\caption{\label{fig:gumpflow} A plot showing the fitting procedure of the GUMP program. The semi-forward ($\xi=0$) fit that consists of four individual fits for each $t$-dependent PDF is performed before the off-forward ($\xi\not = 0$) fit. The results of the semi-forward fit are then passed to the off-forward fit as fixed parameters, whereas the off-forward parameters will be determined with the off-forward constraints.}
\end{figure}

As discussed before, the whole fit will be split into the semi-forward ($\xi =0$) part and the off-forward ($\xi \not =0$) part to avoiding dealing with the huge parameter set in a single fit. Furthermore, in the semi-forward case constraints on different species $H$, $E$, $\widetilde{H}$, and $\widetilde{E}$ decouple, unlike the off-forward case where all four of them are involved. This allows one to further decompose the semi-forward fit into four separate fits that do not interfere for each of the four $t$-dependent PDFs. Therefore, the whole fitting procedure eventually consists of five individual fits as shown in figure \ref{fig:gumpflow}. Correspondingly, the total $\chi^2$ can be decomposed into five parts:
\begin{equation}
    \chi^2_{\text{tot}} = \chi^2_{\text{fwd}} + \chi^2_{\text{off-fwd}} = \chi^2_{H}+\chi^2_{E}+\chi^2_{\widetilde{H}}+\chi^2_{\widetilde{E}}+ \chi^2_{\text{off-fwd}}\ ,
\end{equation}
where each term $\chi^2_{F}$ with $F= \{H,E,\widetilde{H},\widetilde{E}\}$ stands for the $\chi^2$ for the semi-forward fits of the corresponding $t$-dependent PDFs. 
Then one just needs to minimize these $\chi^2$s separately in order to perform the fit and the results are summarized in table \ref{table:chi2summary}. More details of the fitted parameters are presented in Appendix \ref{app:gumpparam}.

\begin{table}[t]
    \centering
    \def\arraystretch{1.25}
    \begin{tabular}{m{3cm} m{2cm} m{2cm} m{2cm}}
    \hline \hline
    Sub-fits & $\chi^2$  & $N_{\text{data}} $ & $\chi^2_\nu\equiv \chi^2/\nu$\\ \hline
    \multicolumn{4}{l}{\textbf{Semi-forward}}\\
    $t$PDF $H$             & 281.7 & 217  & 1.41  \\
    $t$PDF $E$             & 59.7  & 50   & 1.36  \\
    $t$PDF $\widetilde{H}$ & 159.3  & 206  & 0.84  \\
    $t$PDF $\widetilde{E}$ & 63.8  & 58   & 1.23  \\ \hline
    \multicolumn{4}{l}{\textbf{Off-forward}} \\
    JLab DVCS              & 1413.7  & 926  & $\sim$ 1.53\\ 
    H1 DVCS                & 19.7  & 24   & $\sim$ 0.82\\ 
    Off-forward total      & 1433  & 950 & \textbf{1.53}\\ \hline 
    \textbf{Total}         & 2042  & 1481 & \textbf{1.40}\\
    \hline\hline
    \end{tabular}
    \caption{A table summarizing the total $\chi^2$ versus the number of data points $N_{\text{data}}$ as well as the $\chi^2$ per degree of freedom $\chi^2_\nu$. Since the JLab and H1 DVCS measurements are fitted simultaneously in the off-forward fit, the $\chi^2/\nu$ is replaced with $\chi^2/N_{\text{data}}$ instead to estimate their separate contributions.}
    \label{table:chi2summary}
\end{table}

Several comments are as follows: first, we take the globally fitted unpolarized and polarized PDFs~\cite{Cocuzza:2022jye} with 31 points sampled in the region $x\in[0.005,0.6]$ for each flavor, indicating 155 points for the $t$-dependent PDFs $H$ and $\widetilde{H}$ respectively. Note that since the parameterization we used has only 3 parameters in the forward limit for each PDF, much less than the PDF global analysis~\cite{Hou:2019efy}, the sample size can not be too large correspondingly. The extra constraints from PDFs on $H$ and $\widetilde{H}$ account for their larger $N_{\rm{data}}$ as shown in table \ref{table:chi2summary}, while the other around 50 data for each $t$-dependent PDF are taken from the globally fitted FFs~\cite{Ye:2017gyb} and lattice calculations~\cite{Alexandrou:2021jok,Alexandrou:2020zbe}.

\begin{figure}[t]
\centering
\includegraphics[width=\textwidth]{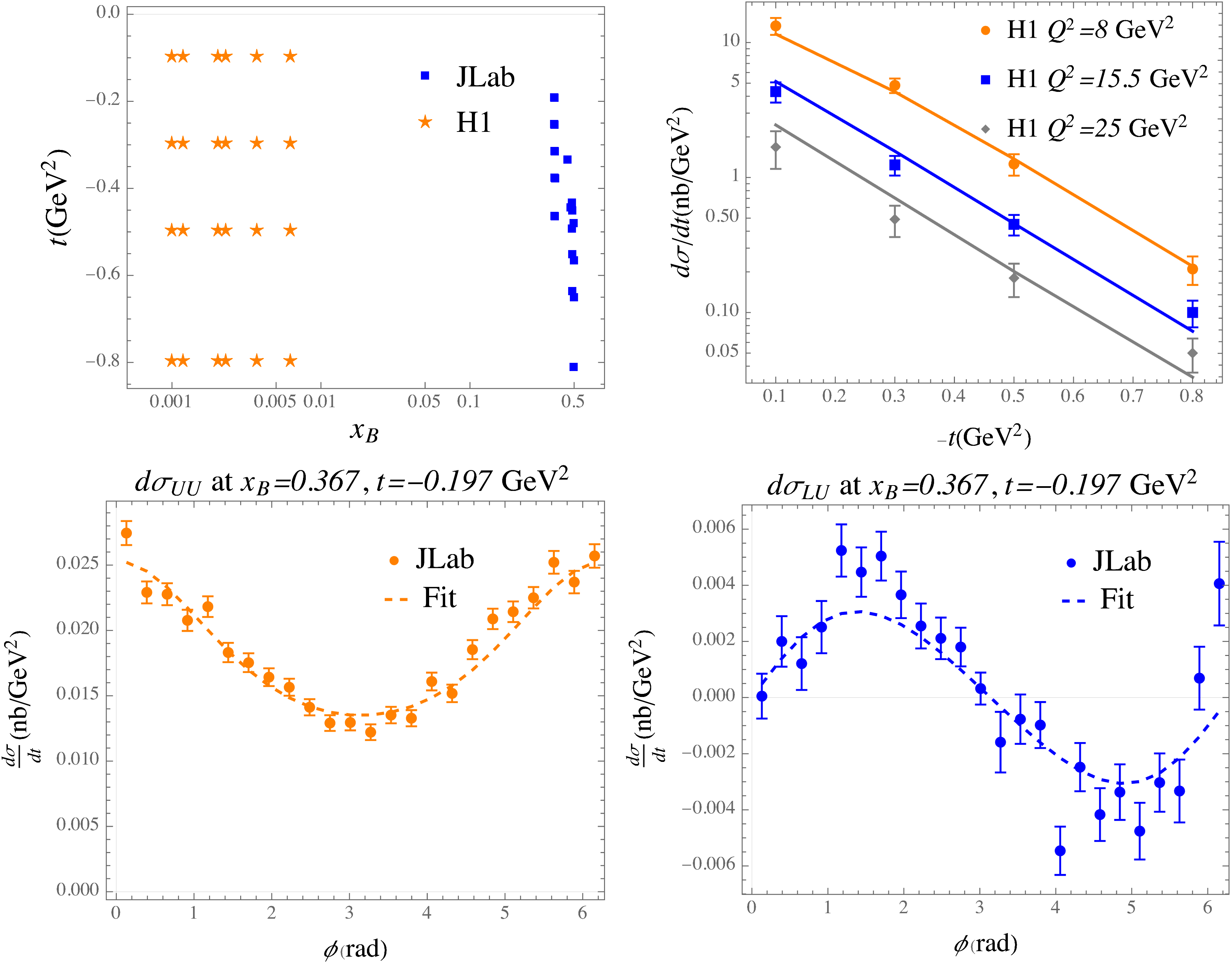}
\caption{\label{fig:DVCSfit} The top left panel shows the kinematical coverage of the DVCS measurements including both JLab and H1. The top-right panel shows the fit to the DVCS measurements at H1 with the azimuthal angel $\phi$ integrated, whereas the lower two figures show examples of the fit to the JLab DVCS cross-section measurements at $Q^2=3.65\text{ GeV}^2$ in the azimuthal angel $\phi$ for both polarized and unpolarized beam. Contributions of Bethe-Heitler process are calculated with the electromagnetic form factors in ref. \cite{Ye:2017gyb}. The DVCS and interference cross-sections are calculated with the formulas in ref. \cite{Guo:2021gru} including the contributions of twist-two CFFs only.}
\end{figure}

Second, in the interest of this work, we select the JLab DVCS data with larger $Q^2$ ($Q>1.8$ GeV) and smaller $x_B$ ($x_B<0.5$). The larger $Q^2$ is required by the factorization theorem and also to suppress the higher twist effects, while the low $x_B$ region is selected in accord with the small $\xi(x_B)$  expansion such that we have the expansion parameter $\xi^2 \lesssim 0.1$. Even with such selection, it still leaves much more JLab DVCS data than the H1 data (which do not need selection as they are already in the large $Q^2$ and small $x_B$ region), since both the azimuthal $\phi$ dependence and beam-polarized cross-sections are measured at JLab. Therefore, the JLab data will form the dominant input in the off-forward fit even though they cover about the same number of kinematical points on the $(x_B,t)$ space. In figures \ref{fig:DVCSfit}, we show the kinematical coverage of the DVCS measurements at both JLab and H1 and present some typical examples of the fit to the DVCS measurements.

To summarize, the reduced $\chi^2$s are all around 1 which indicates generally good fits. For the forward fit, the main issue is that the 3 forward parameters for the $H$ PDF cannot perfectly describe the $x$-shape ranging from $x\in[0.005,0.6]$. Although this could be improved with more sets of model ansatz, it would also require more off-forward inputs accordingly. As for the off-forward fit, the main challenge seems to be the potential higher-twist effects in the JLab DVCS measurements given that the typical $Q^2$ of them is around 4 GeV$^2$, which still allows sizeable twist-three contributions or even twist-four contributions especially at large momentum transfer~\cite{Guo:2022cgq,Braun:2014sta,JeffersonLabHallA:2022pnx}. For instance, the LU cross-section plot at the lower-right of figure \ref{fig:DVCSfit} still deviates from a perfect sine shape as predicted in the $Q^2\to \infty$ limit, which could be due to the higher-twist effects.

\subsection{The extraction of Compton form factors}

\begin{figure}[t]
\centering
\includegraphics[width=1\textwidth]{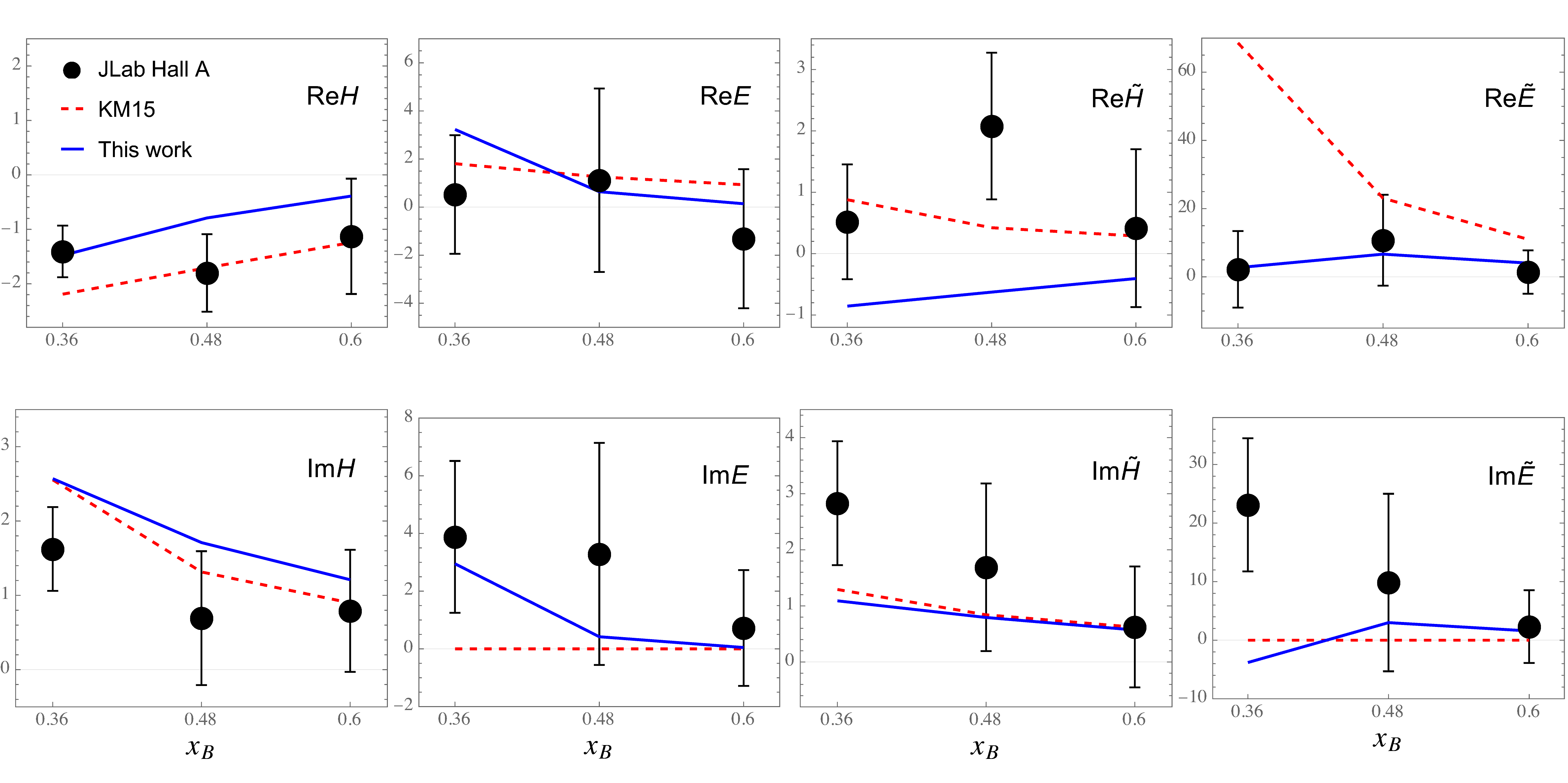}
\caption{\label{fig:CFFcompare} A comparison of the CFFs extracted in this work with the locally extracted CFFs in ref.~\cite{JeffersonLabHallA:2022pnx} as well as the CFFs predicted by the KM15 model~\cite{Kumericki:2015lhb}. The three different kinematical points with $x_B = 0.36,0.48,0.60$ have momentum transfer $t = -0.345, -0.702,-1.050$ GeV$^2$ respectively. The CFFs in ref.~\cite{JeffersonLabHallA:2022pnx} are extracted from data with various different $Q^2$, whereas the theoretical values are calculated at the reference scale $Q^2  = 4$ GeV$^2$.}
\end{figure}

As discussed before, since the off-forward inputs contain DVCS measurements only which are effectively combinations of CFFs, they do not constrain the $x$ shape nor the flavor structures of GPDs well. Thus, the flavor-dependent $x$ shape of GPDs extracted under the empirical constraints in subsection \ref{subsec:momentpara} will be obviously model-dependent. Therefore, before moving on to the extracted GPDs which relies on the choice of ansatz, we first discuss the extracted CFFs of which the comparison is less model-dependent. More specifically, we will compare the CFFs extracted in this work with the locally extracted CFFs in ref.~\cite{JeffersonLabHallA:2022pnx} and the CFFs predicted by the KM15 model~\cite{Kumericki:2015lhb} based on the global analysis of CFFs. The comparison of the CFFs are shown in figure \ref{fig:CFFcompare}.

Before commenting on the comparison of the extracted CFFs, we first note that even the local extraction of CFFs suffers from the degeneracy issue. The total DVCS cross-sections are generally quadratic equations of all the 8 CFFs:
\begin{equation}
        \text{d}\sigma^{\rm{P_b P_t}} = \text{d}\sigma^{\rm{P_b P_t}}_{\rm{DVCS}} + \text{d}\sigma^{\rm{P_b P_t}}_{\rm{INT}} + \text{d}\sigma^{\rm{P_b P_t}}_{\rm{BH}} = \sum_{i,j} A^{\rm{P_b P_t}}_{ij}\mathcal{F}^i \mathcal F^{j} + \sum_{i} B^{\rm{P_b P_t}}_{i}\mathcal{F}^i  + C^{\rm{P_b P_t}}\ ,
\end{equation}
with $\mathcal F_{i}=\{\text{Re}\mathcal{H},\text{Im}\mathcal{H},\text{Re}\mathcal{E},\text{Im}\mathcal{E},\text{Re}\mathcal{\widetilde{H}},\text{Im}\mathcal{\widetilde{H}},\text{Re}\mathcal{\widetilde{E}},\text{Im}\mathcal{\widetilde{E}}\}$ which are the real or imaginary parts of the CFFs corresponding to the GPDs $H$, $E$, $\widetilde{H}$ and $\widetilde{E}$.
For each combination of beam polarization $P_b$ and target polarization $P_t$, the pure DVCS (DVCS), interference (INT) and Bethe-Heitler (BH) contributions are quadratic, linear, and constant in the CFFs respectively.
Ideally, one would need all 8 possible combinations of $P_b$ and $P_t$ in order to disentangle the 8 CFFs, see for instance ref. \cite{Shiells:2021xqo}, but there could still be degeneracy in the solutions as the nature of quadratic equations.

In this work, the degeneracy will be more severe, since only two polarization configurations, unpolarized or polarized beam with unpolarized target (UU and LU), are considered. For instance, one can show that with UU and LU cross-section, the CFF $\widetilde{\mathcal{E}}$ only shows in the quadratic terms, multiplied to either itself or $\widetilde{\mathcal{H}}$, implying that the quadratic terms are invariant under the transformation
\begin{equation}
    \text{Re}\widetilde{\mathcal{H}} \to - \text{Re}\widetilde{\mathcal{H}} \quad ,\text{Re}\widetilde{\mathcal{E}} \to - \text{Re}\widetilde{\mathcal{E}}\ ,
\end{equation}
and the same for the imaginary part. This degeneracy of CFFs in the DVCS cross-sections leads to the ambiguity in the sign of the extracted $\text{Re}\widetilde{\mathcal{H}}$ and $\text{Re}\widetilde{\mathcal{E}}$ as well as the $\text{Im}\widetilde{\mathcal{H}}$ and $\text{Im}\widetilde{\mathcal{E}}$ shown on the right of figure \ref{fig:CFFcompare}, where the extracted CFFs in this work seem to take opposite sign compared to that of the local extraction.\footnote{We note that when testing the fitting program with slightly different set-up, the extracted CFFs indeed turn out to have different signs.} Besides this explicit example, there might be other implicit degeneracies in the extracted CFFs which could affect the reliability of such an extraction.

Although such degeneracy makes the comparison of the extracted CFFs more subtle and less intuitive, it can certainly be improved in the future with more polarization configurations taken into consideration. On the other hand, many of the CFFs extracted in this work agree well with the local extraction as shown in figure \ref{fig:CFFcompare}, adding more confidence to the extraction of those CFFs.

\subsection{The extracted GPDs at non-zero skewness}

Compared to the CFFs extraction discussed in the previous subsection, the extraction of GPDs will involve more model dependence. The lack of off-forward constraints except the CFFs is the main challenge in the GPD extraction, and the extracted $x$-shape of GPDs will depend on the ansatz chosen correspondingly. In addition, as the CFFs are averaged over different flavors, the extracted flavor structures could be ambiguous as well. Though suffering from these ambiguities, we still present the extracted GPDs here as an illustration of how GPDs are constrained by the inputs, with the caveat that the GPDs, especially their off-forward behaviors, are not uniquely determined at this point. These results can certainly be improved with more lattice calculation at non-zero skewness as well as more flavor-sensitive data in the future.

\begin{figure}[t]
\centering
\includegraphics[width=\textwidth]{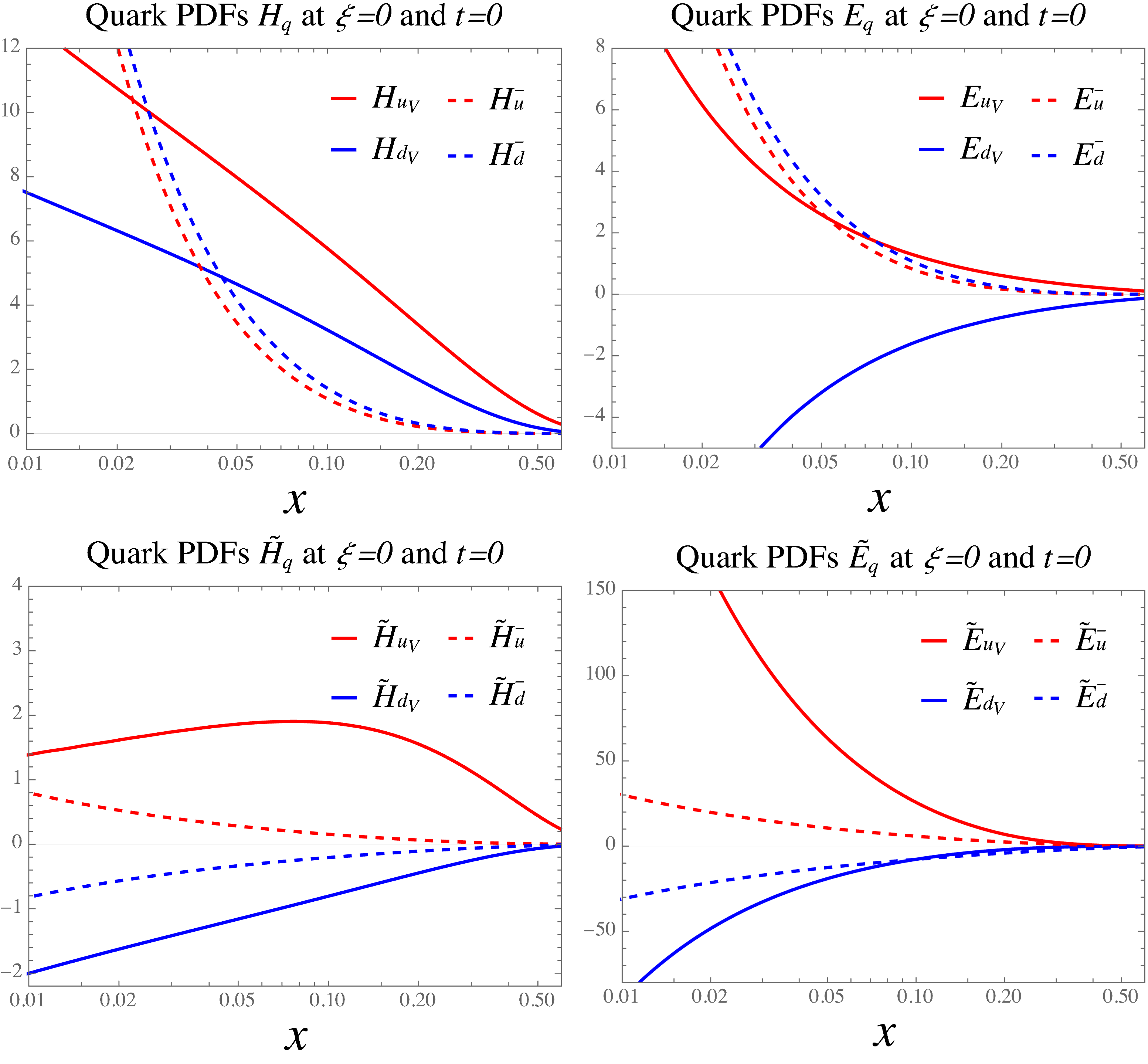}
\caption{\label{fig:gumppdf}Plots present some examples of the extracted quark PDFs at $\xi=0$, $t=0$ and the reference scale $\mu_0= 2$ GeV. The $H$ and $\widetilde{H}$ are obtained from the polarized and unpolarized PDFs whereas the $E$ and $\widetilde{E}$ are constrained by the form factors and lattice calculation of form factors as well as GPDs. The gluon PDFs are also implemented in this work, and they enter the scale evolutions and mix with the quark GPDs. However, since they are not as constrained by the DVCS measurements, they are not presented here.}
\end{figure}

In figure \ref{fig:gumppdf}, we present the extracted PDFs $H$, $E$ and $\widetilde{H}$, and $\widetilde{E}$ at $\xi=0$, $t=0$, and the reference scale $\mu_0 = 2$ GeV. The two PDFs $H$ and $\widetilde{H}$ are fully parameterized and fitted to the globally extracted unpolarized and polarized PDFs in ref. \cite{Cocuzza:2022jye}, and therefore they agree well with the reference values there, which are not shown in the plots though. On the other hand, due to the lack of the information, the PDFs $E$ and $\widetilde{E}$ are extracted with the empirical constraints summarized in table \ref{table:GUMPparameters} with globally extracted FFs and lattice calculation of form factors as well as GPDs. The $\widetilde{E}$ PDFs turn out to be quite significant due to the contributions of the pion pole, according to the lattice calculated form factors of $\widetilde{E}$~\cite{Alexandrou:2021jok} that we fit the $\widetilde{E}$ PDF to. As for the $E$ PDFs, the shape of the $E$ PDFs are obtained combining the lattice calculation of $E$ GPDs and other relevant form factors from both experiments and lattice results. The $u$ and $d$ quark $E$ GPDs are almost the opposite of each other, in accord with the observation that the flavor isoscalar form factors $B_{u+d}$ and gluonic gravitational form factors $B_{g}$ are consistent with zero according to the lattice results~\cite{Hagler:2009ni,Alexandrou:2021jok,Pefkou:2021fni}. We note that the valence part of the above PDFs are obtained in the semi-forward fit of the previous work~\cite{Guo:2022upw} as well. However, they are quantitatively slightly different from the previous results because of the different constraints used here as well as the effects of the sea quark distributions that were not considered in the previous work.

\begin{figure}[t]
\centering
\includegraphics[width=\textwidth]{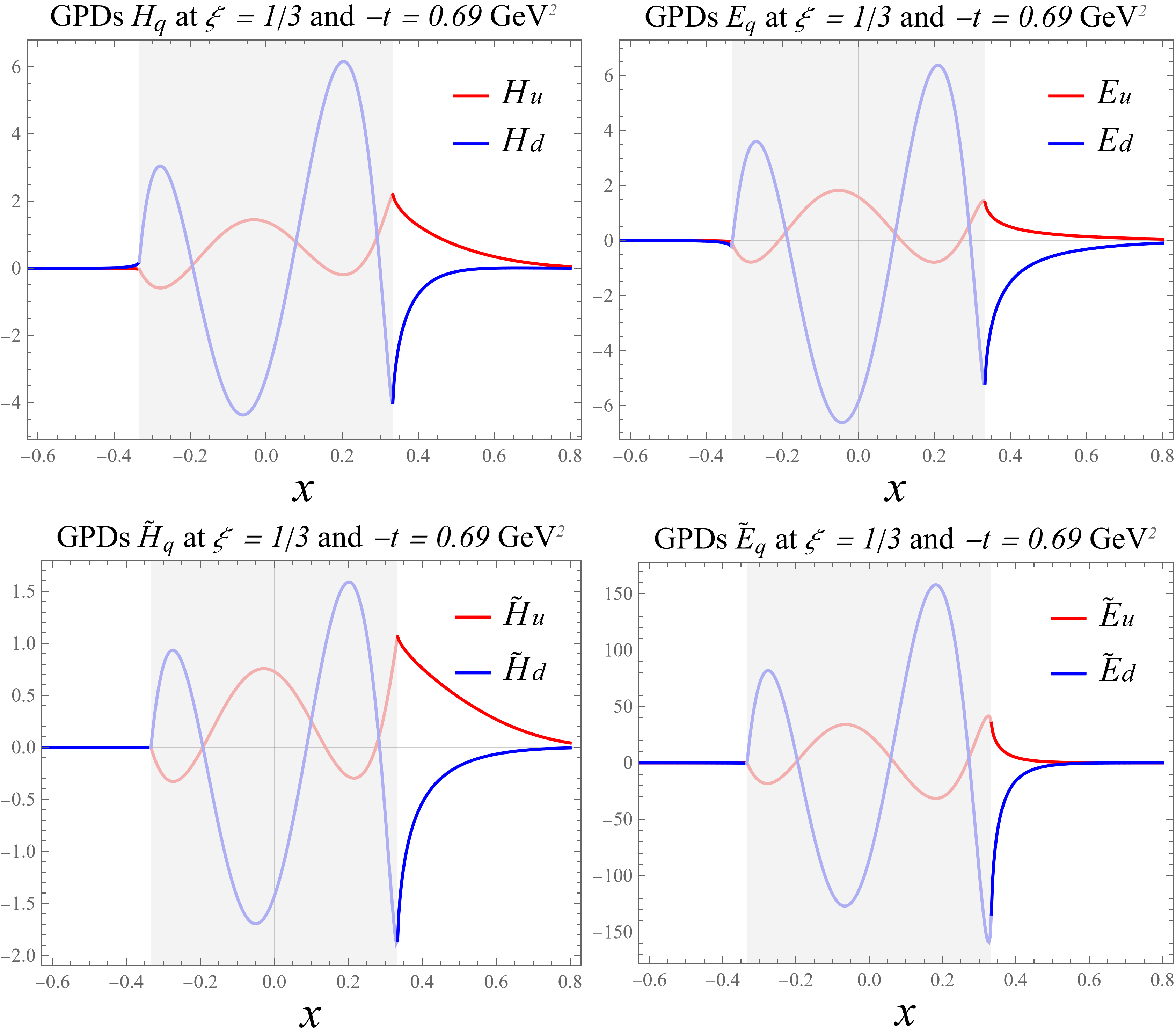}
\caption{\label{fig:gumpgpd} Plots of quark GPDs at $\xi=1/3$, $t=-0.69\text{ GeV}^2$ and reference scale $\mu_0 =2$ GeV. The DA-like regions of the GPDs ($-\xi<x<\xi$) are shaded, and the curves there are plotted in lighter color for distinction. The oscillating behaviors of the GPDs in the DA-like region are the results of the conformal partial wave expansion, since the GPDs here are expanded in terms of the Gegenbauer polynomials there. We note that the shape of GPDs are determined by the model choice of the moments and the CFFs effectively constrain the GPDs at the crossover line $x=\pm\xi$ only to the leading order.}
\end{figure}

In figure \ref{fig:gumpgpd}, we present quark GPDs at $\xi=1/3$ and $t=-0.69$ GeV$^2$ and reference scale $\mu_0 =2$ GeV. In general, the extracted GPDs oscillate in the DA-like region ($-\xi<x<\xi$) with a damping tail in the PDF-like region ($x>\xi$ or $x<-\xi$), as the result of the conformal partial wave expansion which expands the GPDs in the DA-like region in terms of Gegenbauer polynomials that oscillate. We note that due to the DA terms of GPDs, namely the $F_{q\bar q}$ terms in eq. (\ref{eq:gpddecomp}), the behavior of GPDs in the DA-like region could look very different from here. With off-forward inputs mainly just CFFs, only the GPDs at the crossover line $x=\pm\xi$ are effectively constrained, which are then extrapolated with decaying tails to the PDF-like region, whereas the DA-like regions are not uniquely determined which we will discuss more in the next subsection.

The above extracted PDFs and GPDs are generated with the open-source codes of this program~\cite{Guo:2022gumpgit} which could be used to generate other observables including DVCS cross-section measurements as well, although we should note that only the central values of them are available at present. While the error propagation is a crucial part of a global analysis, the process is extremely computationally intensive. With 20 functions of three variables where each point must be calculated through a numerical contour integral, sampling them over more than 50 parameters for the statistical uncertainties would be very challenging though not as useful since the main uncertainties are still from the systematics. Therefore, we will leave the error estimation to the future works once the task will be more practical.

\subsection{GPDs in the DA-like region and DA terms}

In the end of this section, we discuss more about the GPDs in the DA-like region.
Recall that in the previous section, we discussed that in the small $x_{B}$ expansion the DA-like region gets less relevant, and so we did not consider the DA-terms $F_{q\bar q}$ in the global analysis. However, the behavior of GPDs in this region is of genuine interest as well, which is less known compared to our knowledge of GPDs in the PDF-like region e.g., from PDFs. This region of GPDs is equally important and can be accessed directly from lattice calculations at non-zero skewness~\cite{Alexandrou:2020zbe}. We note that these constraints on GPDs in the DA-like regions were not imposed in the global analysis since only very few of them are available at present. However, in this subsection, we will discuss how the current framework under small $x_{B}$ expansion can be extended to accommodate these constraints with the extra DA-terms $F_{q\bar q}$ and produce smoother GPDs in the DA-like region.

In eq. (\ref{eq:conformalsum}), we showed that GPDs can be expressed as the sum of their conformal partial wave, where each term is given by a rescaled Gegenbauer polynomial~\cite{Mueller:2005ed}:
\begin{equation}
  (-1)^j p_j(x,\xi)\equiv  \xi^{-j-1} \frac{2^j \Gamma\left(\frac{5}{2}+j\right)}{\Gamma{\left(\frac{3}{2}\right)}\Gamma(j+3)} \left[1-\left(\frac{x}{\xi}\right)^2\right] C_{j}^{\frac{3}{2}}\left(\frac{x}{\xi}\right)\quad \text{for }|x|<\xi\ ,
\end{equation}
that is non-zero only the in DA-like region. Therefore, one can always add finite terms like this to the GPDs freely without changing the GPDs in the PDF-like region. Such terms are called the DA terms in this work. Since the Wilson coefficients are antisymmetric in $x$ for the CFFs $\mathcal{H}$ and $\mathcal{E}$, adding terms with even $j$ which are symmetric in $x$ will keep both the CFFs $\mathcal{H}$ and $\mathcal{E}$ invariant, which correspond to the so-called shadow GPDs~\cite{Bertone:2021yyz}. On the other hand, adding terms with odd $j$ which are antisymmetric in $x$ will modify the real part of the CFFs $\mathcal{H}$ and $\mathcal{E}$ while keeping their imaginary parts the same, which are also known as the subtraction terms~\cite{Kumericki:2009uq}. Similar arguments apply to the $\widetilde{\mathcal{H}}$ and $\widetilde{\mathcal{E}}$ CFFs too, except that their Wilson coefficients are symmetric in $x$.

\begin{figure}[t]
\centering
\includegraphics[width=0.8\textwidth]{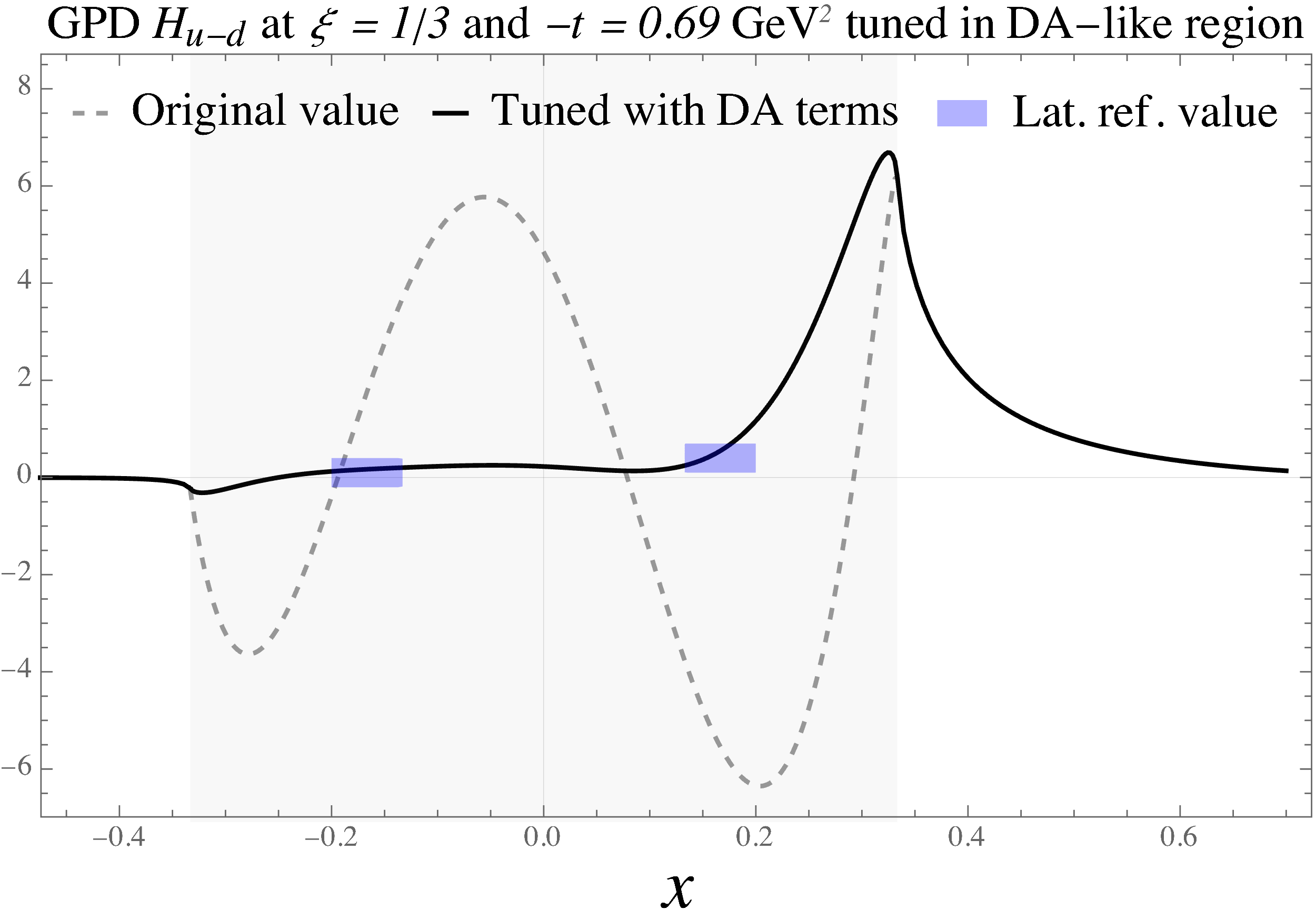}
\caption{\label{fig:gumpDAcomp}The isovector GPD $H_{u-d}$ tuned to fit the reference value calculated on lattice~\cite{Alexandrou:2020zbe} at reference scale $\mu_0=2$ GeV. The dashed line is the original extracted shape of GPDs. By adding extra terms in the DA-like region, we obtain the tuned GPDs as the red curve which approach the reference value shown as blue blocks.}
\end{figure}

These terms are in principle hard to extract from experiments, however, they do affect the generalized form factors and can be constrained by the FFs as well as lattice calculations. Therefore, one should parameterize these extra DA terms, at least part of them, and fit them to these constraints in order to obtain the shape of GPDs in the DA-like region.
There have been lattice calculations of the GPD shape~\cite{Alexandrou:2020zbe}, which could be used to constrain such terms and determine the shape of GPDs in the DA-regions. However, since the results contain only the isovector $u-d$ combination of $H$ and $\widetilde{H}$ GPDs and such calculations will break down at $x=\pm \xi$, they do not pose enough constraints on the GPDs for global analysis. Therefore, we will leave the extra fitting of the DA terms to those constraints in the future work with more information in the DA-like region. In figure \ref{fig:gumpDAcomp}, we show an example of how GPDs can be tuned with the extra DA terms to fit to the constraints in the DA-like regions, where we take the isovector GPDs $H_{u-d}$ calculated on lattice~\cite{Alexandrou:2020zbe} as the reference value. 

\begin{figure}[t]
\centering
\includegraphics[width=0.8\textwidth]{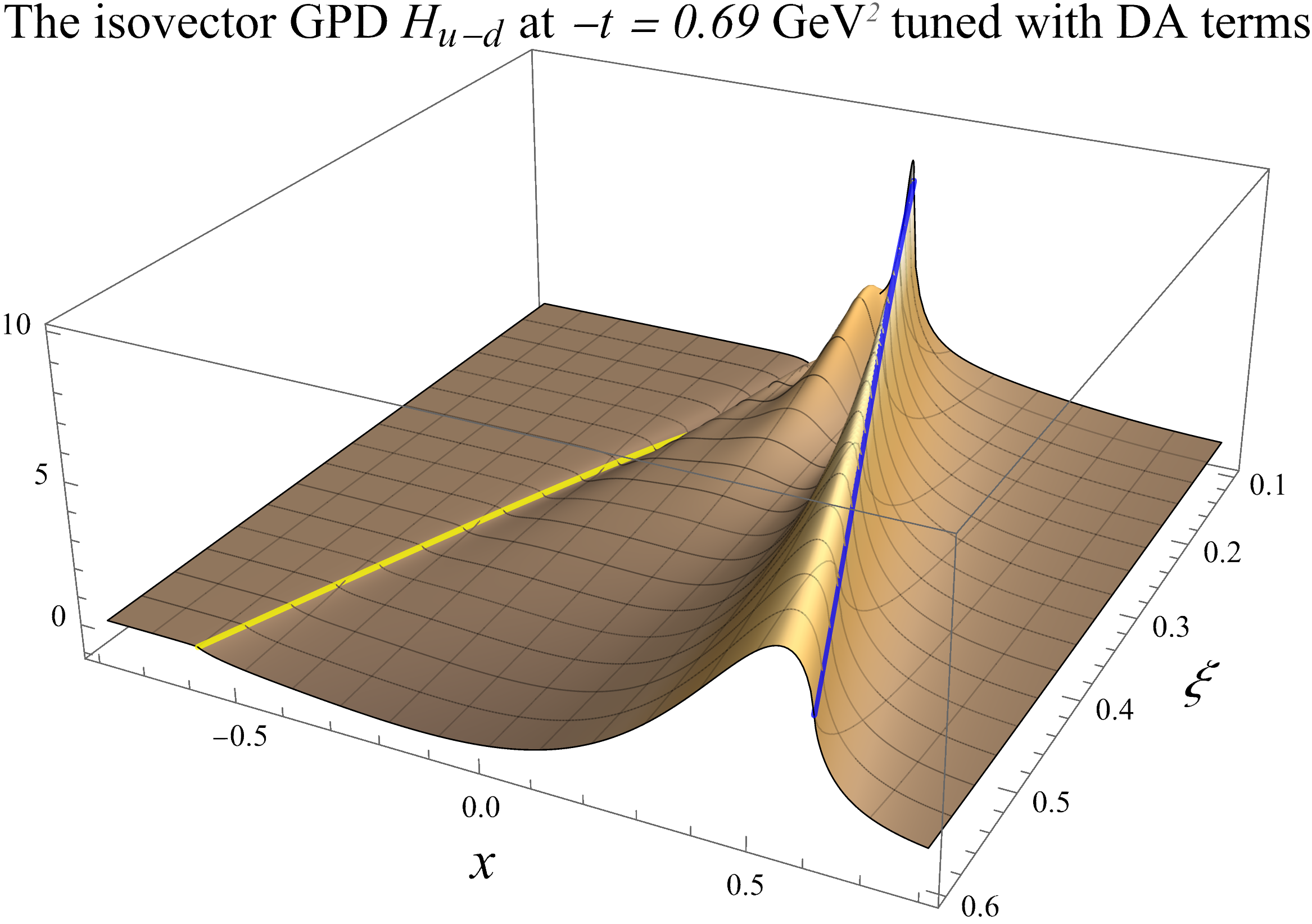}
\caption{\label{fig:gump3d}The isovector GPD $H_{u-d}$ on the $(x,\xi)$ plane tuned to fit the reference value calculated on lattice~\cite{Alexandrou:2020zbe} at $t=-0.69$ GeV$^2$ and reference scale $\mu_0=2$ GeV. The blue ($x=\xi$) and yellow ($x=-\xi$) curves correspond to GPDs on the two crossover lines respectively. A cut at $\xi=0.1$ is made since GPDs in the DA-like region $-\xi<x<\xi$ get singular when $\xi$ approaches $0$.}
\end{figure}

In figure \ref{fig:gump3d}, we also show the tuned isovector GPD $H_{u-d}$ on the $(x,\xi)$ plane at $t=-0.69$ GeV$^2$ and reference scale $\mu_0=2$ GeV. We again note that such results are obtained under the ansatz and empirical constraints used in this work. 

\section{Conclusion and outlook}
\label{sec:conclude}

We extend the previous work~\cite{Guo:2022upw} of the GUMP program to the non-zero skewness case and perform the global analysis of quark GPDs combing experimental measurements of DVCS, relevant lattice calculations for GPDs and PDFs and FFs from global analysis for the first time, whereas the gluon GPDs will be carried out in a separate work with other gluon-sensitive processes such as DVMP.

We argue that empirical constraints are still needed for the global analysis of GPDs, given the extremely large system of GPDs that one needs to consider and the limited knowledge of them currently. With these empirical constraints, we extract the GPDs from global analysis with the above inputs and present the globally extracted PDFs, CFFs and GPDs, with the caveat that more inputs, including more polarization configurations for the DVCS measurements and more lattice results of GPDs at non-zero skewness, are still needed to improve the reliability of such an extraction.

We also discuss the general framework to extend the current program which focuses on the small $\xi$ region of GPDs to allow the analysis of the DA-like regions of GPDs that will be more relevant at larger $\xi$. We argue that besides the quark and antiquark GPDs, the extra DA terms are crucial in describing the GPDs in the DA-like regions, which can improve the parameterization with more flexibility and without damaging the physical constraints like polynomiality conditions. We present an example how the DA terms could modify the GPDs in the DA-like region while keeping the CFFs the same, and therefore can be used to parameterize and fit the shape of GPDs in this region.

In the future works, we will consider the meson production processes in the global analysis to better constrain the gluon GPDs. In addition, we also consider adding the strange quark distributions to the analysis which might have sizable effects. Besides, we will also include the DA terms in the global analysis to fit the GPDs in the DA-like region once enough constraints on the GPDs in the DA-like region are obtained.

\section*{Acknowledgments}

We thank K. Kumeri\v{c}ki for discussions related to the subject of this paper as well as the Gepard package~\cite{gepard} for the useful open-source codes. This research is supported by the U.S. Department of Energy, Office of Science, Office of Nuclear Physics, under contract number DE-SC0020682, and the Center for Nuclear Femtography, Southeastern Universities Research Association, Washington D.C. This research is also supported by the 3D quark-gluon structure of hadrons: mass, spin, and tomography (QGT) topical collaboration.

\appendix

\section{GUMP parameters and their best-fit values}

\label{app:gumpparam}
In this appendix we present more details of the fit, especially the best-fit parameters obtained from the global analysis. In table \ref{table:gumpparam}, we show all the independent parameters with their statistic uncertainties estimated by the Hessian matrix with the \textit{Minuit2} package.

\begin{table}[b!]
    \def\arraystretch{1.25}
    \centering
    \begin{tabular}{|c | c |c| c|}
    \hline
    \multicolumn{2}{|c|}{Vector GPDs $H$ and $E$} & \multicolumn{2}{c|}{Axial-vector GPDs $\widetilde{H}$ and $\widetilde{E}$} \\ \hline
    Parameter  & Value (uncertainty) & Parameter  & Value (uncertainty) \\ \hline
    $N^{H}_{u_V}$ & 4.923 (89) & $N^{\widetilde{H}}_{u_V}$ & 4.833 (429)\\ \hline
    $\alpha^{H}_{u_V}$ & 0.216 (7) & $\alpha^{\widetilde{H}}_{u_V}$ & -0.264 (34)\\ \hline
    $\beta^{H}_{u_V}$ & 3.229 (23) & $\beta^{\widetilde{H}}_{u_V}$ & 3.186 (122)\\ \hline
    $\alpha'^{H}_{u_V}$ & 2.347 (51)& $\alpha'^{\widetilde{H}}_{u_V}$ & 2.182 (175)\\ \hline
    $N^{H}_{\bar u}$ & 0.163 (8)& $N^{\widetilde{H}}_{\bar u}$ &0.070 (33)\\ \hline
    $\alpha^{H}_{\bar u}$ &1.136 (10)  & $\alpha^{\widetilde{H}}_{\bar u}$ & 0.538 (112)\\ \hline
    $\beta^{H}_{\bar u}$ & 6.894 (207)& $\beta^{\widetilde{H}}_{\bar u}$ & 4.229 (1320)\\ \hline
    $N^{H}_{d_V}$ & 3.359 (170)& $N^{\widetilde{H}}_{d_V}$ & -0.664 (170) \\ \hline
    $\alpha^{H}_{d_V}$ & 0.184 (18) & $\alpha^{\widetilde{H}}_{d_V}$ & 0.248 (76) \\ \hline
    $\beta^{H}_{d_V}$ & 4.418 (77)& $\beta^{\widetilde{H}}_{d_V}$ & 3.572 (477) \\ \hline
    $\alpha'^{H}_{d_V}$ &3.482 (171)& $\alpha'^{\widetilde{H}}_{d_V}$ & 0.542 (103)\\ \hline
    $N^{H}_{\bar d}$ & 0.249 (12)& $N^{\widetilde{H}}_{\bar d}$ & -0.086 (42)\\ \hline
    $\alpha^{H}_{\bar d}$ & 1.052 (10)& $\alpha^{\widetilde{H}}_{\bar d}$ & 0.495 (137)\\ \hline
    $\beta^{H}_{\bar d}$ & 6.554 (216)& $\beta^{\widetilde{H}}_{\bar d}$ & 2.554 (897)\\ \hline
    $N^{H}_{g}$ & 2.864 (108)& $N^{\widetilde{H}}_{g}$ & 0.243 (304)\\ \hline
    $\alpha^{H}_{g}$ &1.052 (8) & $\alpha^{\widetilde{H}}_{g}$ & 0.631 (330)\\ \hline
    $\beta^{H}_{g}$ & 7.413 (165)& $\beta^{\widetilde{H}}_{g}$ & 2.717 (2865)\\ \hline
    $N^{E}_{u_V}$ & 0.181 (38)& $N^{\widetilde{E}}_{u_V}$ & 7.993 (3480)\\ \hline
    $\alpha^{E}_{u_V}$ &0.907 (17) & $\alpha^{\widetilde{E}}_{u_V}$ & 0.800 (116)\\ \hline
    $\beta^{E}_{u_V}$ & 1.102 (245)& $\beta^{\widetilde{E}}_{u_V}$ &6.415 (1577)\\ \hline
    $\alpha'^{E}_{u_V}$ & 0.461 (86)& $\alpha'^{\widetilde{E}}_{u_V}$ & 2.076 (933)\\ \hline
    $N^{E}_{d_V}$ &-0.223 (47) & $N^{\widetilde{E}}_{d_V}$ & -2.407 (1239)\\ \hline
    $R^{E}_{\rm{sea}}$ & 0.768 (169)& $R^{\widetilde{E}}_{\rm{sea}}$ & 38 (8)\\ \hline
    $R^{H}_{u,2}$ &0.229 (0.032)& $R^{\widetilde{H}}_{u,2}$ & 0.246 (81)\\ \hline
    $R^{H}_{d,2}$ &-2.639 (202)& $R^{\widetilde{H}}_{d,2}$ & 1.656 (375)\\ \hline
    $R^{E}_{u,2}$ & 0.799 (285)& $R^{\widetilde{E}}_{u,2}$ & 2.684 (171)\\ \hline
    $R^{E}_{d,2}$ & 3.404 (1157)& $R^{\widetilde{E}}_{d,2}$ & 38 (2) \\ \hline
    $b^{H}_{\rm{sea}}$ &3.448 (133) & $b^{\widetilde{H}}_{\rm{sea}}$ & 9.852 (1330)\\ \hline
    \end{tabular}
    \caption{\label{table:gumpparam} A summary of the obtained independent GUMP parameters. }
\end{table}

We note that the $N,\alpha,\beta,\alpha',b$ are the parameters in the moments of GPDs according to eq. (\ref{eq:gumpform}) where $\alpha(t)\equiv \alpha + \alpha' t$ corresponds to the linear Regge trajectory and $\beta(t)=\exp(b t)$ corresponds to the extra exponential term. Each of these parameters have a superscript representing its GPD species ($H$, $E$, $\widetilde{H}$ or $\widetilde{E}$) and a subscript representing its flavor ($u_V$, $\bar u$, $d_V$, $\bar d$ or $g$). The $R^{E}_{\rm{sea}}$ and $R^{\widetilde{E}}_{\rm{sea}}$ are the ratio of the $E(\widetilde{E})$ GPDs to the $H(\widetilde{H})$ GPDs for the sea quarks and gluons. Besides them, there are also parameters like $R^{H}_{u,2}$ that are the ratio of the $\xi^2$ terms to the forward terms as defined in eq. (\ref{eq:Rdef}) for GPDs with different species and flavors.  We again note that more details of them and the GUMP program that generate the GUMP GPDs, CFFs and cross-sections are available online~\cite{Guo:2022gumpgit}.

\bibliographystyle{jhep}

\bibliography{refs.bib}

\end{document}